# Superlubricity – a new perspective on an established paradigm

*Oded Hod*

Department of Chemical Physics, School of Chemistry, the Sackler Faculty of Exact Sciences, Tel-Aviv University, Tel-Aviv 69978, Israel

odedhod@tau.ac.il

ABSTRACT: Superlubricity is a frictionless tribological state sometimes occurring in nanoscale material junctions. It is often associated with incommensurate surface lattice structures appearing at the interface. Here, by using the recently introduced registry index concept which quantifies the registry mismatch in layered materials, we prove the existence of a direct relation between interlayer commensurability and wearless friction in layered materials. We show that our simple and intuitive model is able to capture, down to fine details, the experimentally measured frictional behavior of a hexagonal graphene flake sliding on-top of the surface of graphite. We further predict that superlubricity is expected to occur in hexagonal boron nitride as well with tribological characteristics very similar to those observed for the graphitic system. The success of our method in predicting experimental results along with its exceptional computational efficiency opens the way for modeling large-scale material interfaces way beyond the reach of standard simulation techniques.

Layered materials such as graphite, hexagonal boron nitride (*h*-BN), and *2H*-molybdenum disulphide and its fullerene derivatives have long been known to serve as excellent solid lubricants.[1-14] This important characteristic has its origin in their nanoscopic anisotropic crystal structure consisting of strong covalent intra-layer bonding and weaker dispersive interlayer interactions.[15] One important consequence of this unique structure is the fact that the layers may slide on top of each other while overcoming relatively small energetic barriers. Recently, the wearless friction between a nanoscale graphene flake and a graphite surface was measured experimentally as a function of the misfit angle between the two surfaces (see Fig. 1).[16-17] Friction forces ranging from moderate to vanishingly small were obtained depending on the degree of commensurability between the lattices of the flake and the extended surface. The ability to achieve such a state of ultra-low friction, often termed a superlubric state,[18-27] is clearly of high interest both from the basic scientific perspective of nanotribology[28-41] and in light of the promising technological opportunities it carries.[9,42-44]

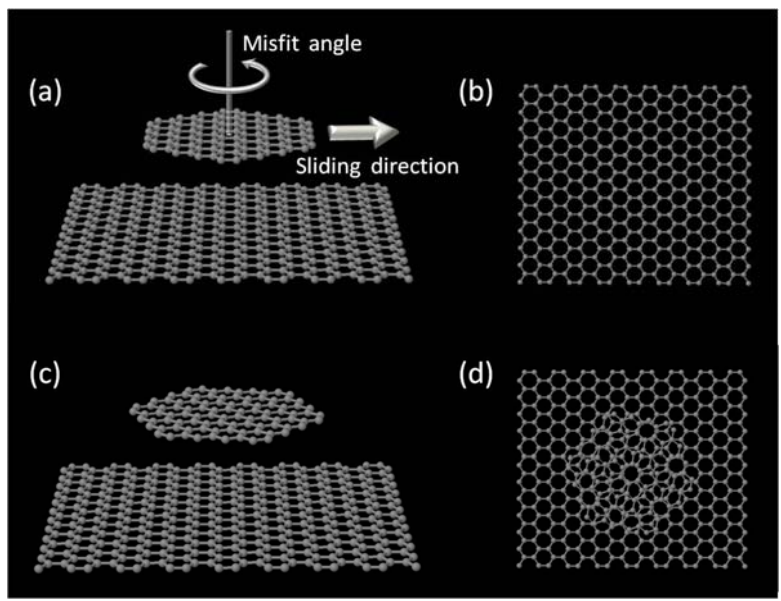

*Fig. 1: Commensurate (panels (a) and (b)) and incommensurate (panels (c) and (d)) configurations of a graphene flake atop of a graphene surface. Both tilted (panels (a) and (c)) and upper (panels (b) and (d)) views of the two arrangements are offered. The definitions of the sliding direction and the misfit angle are indicated in panel (a). The misfit angles for the commensurate and incommensurate modes are $0^o$ and $20^o$, respectively. Clear Moiré patterns appear in panel (d) for the incommensurate state.*

A simple picture that may offer intuitive insights regarding the origin of this relation between friction and commensurability[45-49] is the relative sliding of two "egg-box" foam sheets which are in contact (see Fig. 2). If the lattices imprinted in both foams are initially placed in a commensurate mode (Fig. 2b) when trying to shear the two sheets all unit cells have to cross the physical barriers simultaneously (Fig. 2c-2f) resulting in a stick slip motion leading to a high friction state. If, on the other hand, the lattices are rotated with respect to each other (Fig. 2g), resulting in an incommensurate state, all unit-cells have to cross much smaller barriers at any point in time leading to considerably reduced resistance towards sliding. In the case of layered materials, the physical barriers of the egg-box lattice model are replaced by sliding energy barriers which are mainly a manifestation of enhanced Pauli repulsions between electron clouds centered around different atomic positions on two adjacent layers as they cross each other during the sliding process.[50-53]

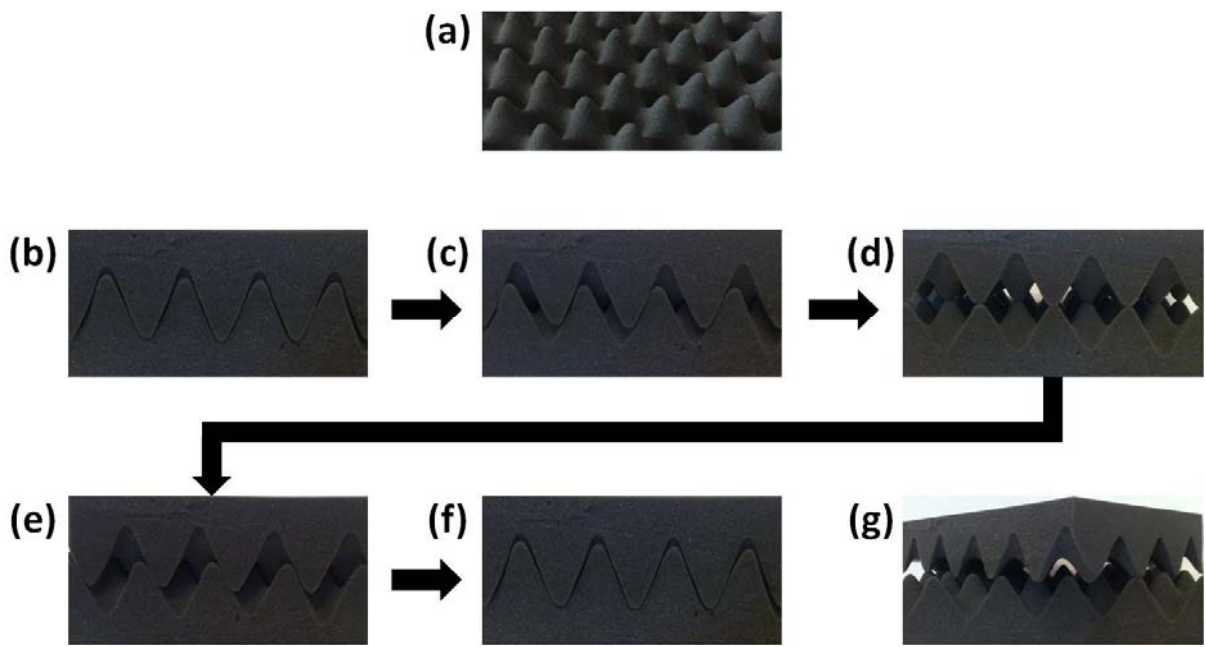

*Fig. 2: Egg-box foam model for commensurate and incommensurate sliding conditions. Panel (a): tilted view of a single egg-box lattice. Panels (b)-(f): relative sliding of two commensurate egg-box foams where all unit cells cross the physical sliding barriers simultaneously. Panel (g): incommensurate orientation of the two lattices.*

More rigorous microscopic understanding of the interlayer sliding process in layered materials has been obtained using sophisticated molecular dynamics simulations,[54-58] semi-empirical approaches,[59] and first-principles[50-51,53,60-66] calculations.[67] Such simulations adopt either a phenomenological[68-74] approach such as the Prandtl[75]-Tomlinson[76] model and its extensions[69] and the Frenkel-Kontorova[77] model or an elaborate atomistic description[49,51-53,61,78-93] of the sliding systems. While such models are often quite successful in reproducing the main physical characteristics of the sliding process in these systems, their degree of sophistication somewhat blurs the direct relation between commensurability and friction. This, in turn, complicates the tasks of identifying the key physical factors responsible for interesting frictional phenomena and designing new materials with novel desired tribological properties.

In this paper we use the recently developed concept of the registry index (RI),[50,94] which gives a quantitative measure of the degree of commensurability between two lattices, to elucidate the observed interplay between interlayer registry and wearless sliding friction in layered materials. As previously speculated, a direct relation between the interlayer registry and the measured interlayer sliding friction is obtained for the case of graphene. Furthermore, fine-tuning of the basic model parameters based on simple intuitive physical considerations results in excellent agreement between the experimental results and the model predictions better than that obtained by sophisticated molecular dynamics calculations and with considerably reduced computational efforts. Finally, we use our validated model to predict the misfit-angle dependence of friction in bilayer hexagonal boron-nitride (*h*-BN).[50]

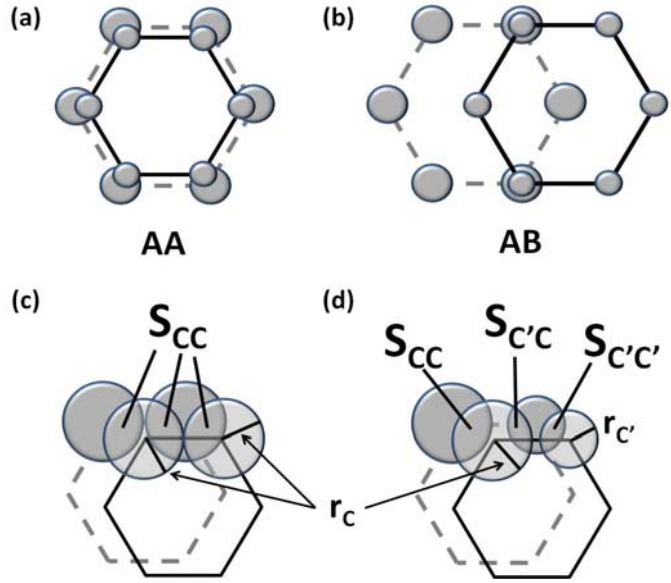

*Fig. 3: **Panel (a)**: worst (AA) stacking mode of a graphene bilayer. **Panel (b)**: optimal (AB) stacking mode of a graphene bilayer. **Panel (c)**: Symmetric registry index definition of the projected overlap area between circles assigned to atomic positions in the upper layer (transparent circles, full lines) and their lower layer counterparts (opaque circles, dashed lines). **Panel (d)**: Asymmetric registry index definition of the projected overlap area between circles assigned to atomic positions in the upper layer (transparent circles, full lines) and their lower layer counterparts (opaque circles, dashed lines). In panels (c) and (d) the circles representing the atomic centers were omitted for clarity.*

The registry index is a simple numerical parameter defined to quantify the interfacial registry mismatch between two lattices. It is defined as a material-dependent function which requires input regarding the surface lattice structures and the optimal and worst (in terms of total energy) inter-surface stacking modes. The general concept may be well delivered by considering the case of an infinite graphene bilayer.[94] Here, the worst stacking mode is the AA configuration where the lattices of the two layers are fully eclipsed (Fig. 3a) and the optimal stacking mode is the AB configuration where half of the carbon atoms in one layer reside atop of the hexagonal centers of the adjacent layer (Fig. 3b). For the definition of the RI each atomic center is assigned with a simple circle of radius $r_C = 0.5 \cdot L_{CC}$, where $L_{CC} = 1.42$ Å is the covalent carbon-carbon bond length within each hexagonal layer (Fig. 3c). Next, the projected overlap between circles belonging to one layer and their counterparts on the adjacent layer is labeled by $S_{CC}$ (Fig. 3c). We note that, similar to the total energy, this overlap obtains a maximum value at the worst (AA) stacking mode and a minimum value at the optimal (AB) interlayer configuration. Noticing that we are looking for a numerical parameter that will quantify the interlayer registry mismatch in accordance with the relative total energies of the different stacking modes we chose our RI to be proportional to the total overlap area $RI \propto S_{CC}$. By normalizing the parameter to the overlap values at the worst ($S_{CC}^{AA}$) and optimal ($S_{CC}^{AB}$) stacking modes in the following manner:

$$RI_{graphitic} = \frac{S_{CC} - S_{CC}^{AB}}{S_{CC}^{AA} - S_{CC}^{AB}}$$

we arrive at a parameter that is bound to the range [0,1] where the value *RI*=1 is obtained for the worst stacking mode and the value *RI*=0 is obtained for the optimal configuration. By plotting the RI for different interlayer configurations of bilayer graphene we have previously shown that the RI landscape captures all important characteristics of the sliding energy landscape obtained by atomistic calculations based on a dispersion augmented tight-binding Hamiltonian.[94]

Having at hand a simple parameter that quantifies the interlayer commensurability at arbitrary stacking modes of bilayer graphene we are now in position to characterize the correlation between experimentally measured friction and the degree of interlayer registry. To this end, we consider the system depicted in Fig. 1 where a finite rigid hexagonal flake slides on-top of an infinite graphene surface. We can define two important parameters: the *misfit angle* and the *sliding direction* (see Fig. 1). The misfit angle is the angle at which the flake is rotated about an axis crossing its center of mass perpendicular to its basal plane such that 0º stands for the orientation of the flake at the AB staking mode. The sliding direction is the direction along which the flake is dragged with respect to the armchair axis of the underlying hexagonal graphene layer.

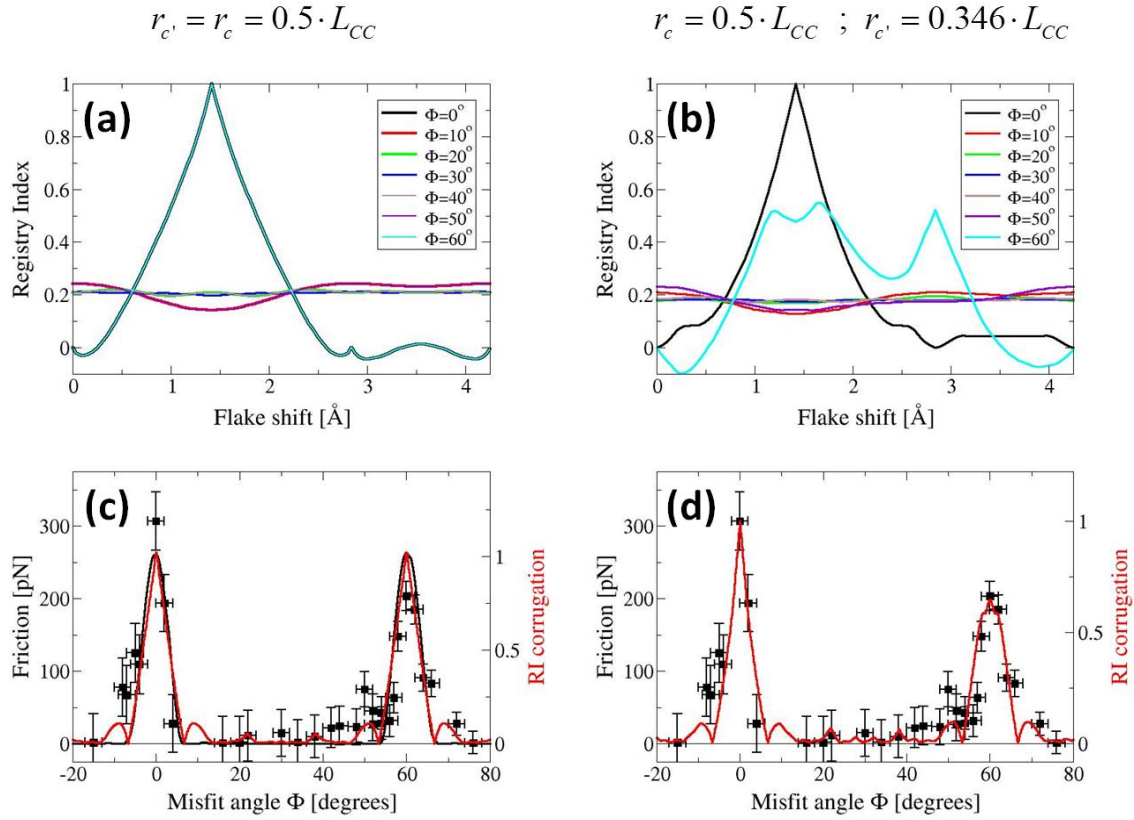

*Fig. 4: Registry index dependence on the misfit angle of a 150 atoms hexagonal graphene flake. Panels (a) and (b): Registry index variations along linear sliding paths at different misfit angles Φ for a single graphene flake on a single graphene sheet (a) and for a multilayer graphene flake on the surface of multilayer graphene (b). Panels (c) and (d): Measured friction (left axis) and corrugation of the registry index landscape along the linear paths (right axis) as a function of the misfit angle for the bilayer (c) and multilayer (d) systems. The sliding direction is chosen to be along the armchair axis of the infinite graphene layer as indicated in Fig. 1(a). Experimental results were reproduced with the kind permission and help of Prof. Joost W.M. Frenken and Prof. Martin Dienwiebel from Dienwiebel, M.; Verhoeven, G. S.; Pradeep, N.; Frenken, J. W. M.; Heimberg, J. A.; Zandbergen, H. W.: Superlubricity of Graphite. Physical Review Letters 2004, 92, 126101. Copyright (2004) by the American Physical Society.*

The RI, now normalized to the overall overlap area of the flake at the AA and AB stacking modes, can be calculated for different misfit angles and sliding directions. In Fig. 4a variations of the registry index along linear paths taken at a sliding direction parallel to the armchair axis of the underlying graphene layer (see Fig. 1a) are presented for different misfit angles. As can be seen, due to the 6-fold symmetry of the system, the registry index behavior for misfit angles of $0°$ and $60°$ is identical presenting large registry variations as the flake slides through the AA and AB stacking modes. For intermediate misfit angles a qualitatively different behavior is obtained where relatively small variations around an average RI value of ~0.2 occur as the flake is dragged along the linear path. Having seen that the RI landscape mimics well the total sliding energy variations in graphene and $h$-BN,[50,94] and assuming that the wearless friction is directly proportional to the sliding energy barriers along a given path, we may plot the corrugation (maximum amplitude) of the RI variations along a given path as a function of the misfit angle and compare the results to the measured friction. This comparison is presented in Fig. 4c where excellent agreement between the RI results and the experimentally measured friction is obtained. Close to the high friction peaks excellent agreement between the RI results and the Tomlinson model predictions (solid black curve) is also obtained. Interestingly, in the low friction region the RI diagram predicts fine features near -9°, 9°, 22°, 38°, 51°, and 69° that are well reproduced by the experimental measurements but are practically absent in the Tomlinson model predictions.

While the agreement achieved between the simple RI theory and experiment is quite impressive, when carefully examining the experimentally measured friction one notices that the high friction peaks are asymmetric while the theoretical results predict symmetric peaks. This was previously attributed to the fact that in experiment both the surface and the flake have a multi-layer structure assumed to possess the ABA stacking mode.[68] Therefore, every two neighboring carbon sites (often marked as C and C') become inequivalent as one resides atop a carbon atom and the other atop a center of a hexagon of the adjacent layer (Fig. 3b). The relevance of this inequivalence to the sliding energy landscape stems from the fact that the π electron clouds are expected to have slightly different shapes and sizes near the C and C' sites. This leads to somewhat altered Pauli repulsions as different sites belonging to the flake and the infinite surface pass against each other throughout the sliding process. As a result the 6-fold symmetry is reduced to a 3-fold symmetry and the high friction peaks at 0° and 60° become asymmetric. To account for this in our RI model we realize that the circles assigned to each atomic cite represent the relative sizes of the atom centered electron clouds and thus we simply assign different radii to the circles associated with the C and C' atomic centers (Fig. 3d). In Fig. 4b variations of the registry index along linear paths taken at a sliding direction parallel to the armchair axis of the underlying graphene layer (see Fig. 1a) are presented for different misfit angles where we have chosen $r_C = 0.5 \cdot L_{CC}$ and $r_{C'} = 0.346 \cdot L_{CC}$. When compared to the results of the symmetric case (Fig. 4a), representing the bilayer system, one immediately realizes that in the multi-layer (asymmetric) case the $\Phi = 0°$ and $\Phi = 60°$ are no longer equivalent and that the overall corrugation of the latter is smaller. For intermediate angles minor difference between the multi-layer and the bilayer representations are obtained where the average corrugation and deviations are very similar in both cases. We note that for some misfit angles negative values of the RI are obtained indicating that for the finite flake at some specific orientations the registry may be better than that obtained at the AB staking mode. Using this data we plot in Fig. 4d the corrugation of the RI variations along a given path as a function of the misfit angle. Now the asymmetry between the $\Phi = 0°$ and $\Phi = 60°$ is fully captured and we obtain a remarkable agreement between the experimentally measured friction and the calculated RI results down to fine details in both high- and low-friction regions.

To further test the validity of the RI model we consider additional aspects of the flake's sliding physics including the dependence of the sliding friction (through the RI corrugation) on the flake size and the sliding direction. In Fig. 5 the dependence of the peak shape on the size of the flake and the sliding direction is presented. For the symmetric (bilayer) case we obtain (Fig. 5a) the expected narrowing of the peak with increased flake size and the relation $tan(\Delta\Phi) = \alpha/D$,[68] where $\Delta\Phi$ is the full width at half-maximum of the peak and D the diameter of the flake expressed in terms of lattice spacings, is fully recovered with $\alpha = 1$ (inset of Fig. 5a). For the asymmetric case (multilayer), a similar peak narrowing is obtained for both peaks (Fig. 5b) though for the higher peak we find that $\alpha < 1$. To better understand this behavior we plot, in Fig. 5c, $tan(\Delta\Phi)$ as a function of $1/D$ for various values of the site asymmetry factor $(r_{C'}/r_C)$ where a monotonic decrease of the slope, $\alpha$, is found as the site asymmetry increases. Interestingly, over a wide range of asymmetry factors a linear relation between the slope and the site asymmetry is obtained. In Fig. 5d we plot the RI corrugation for various sliding directions. As can be seen, at the high friction regions, the sliding direction influences both the absolute and the relative peak heights where in some sliding directions the peak that was originally higher becomes the lower one. For the low friction regions we find merely a marginal effect of the sliding directions considered on the RI corrugation.

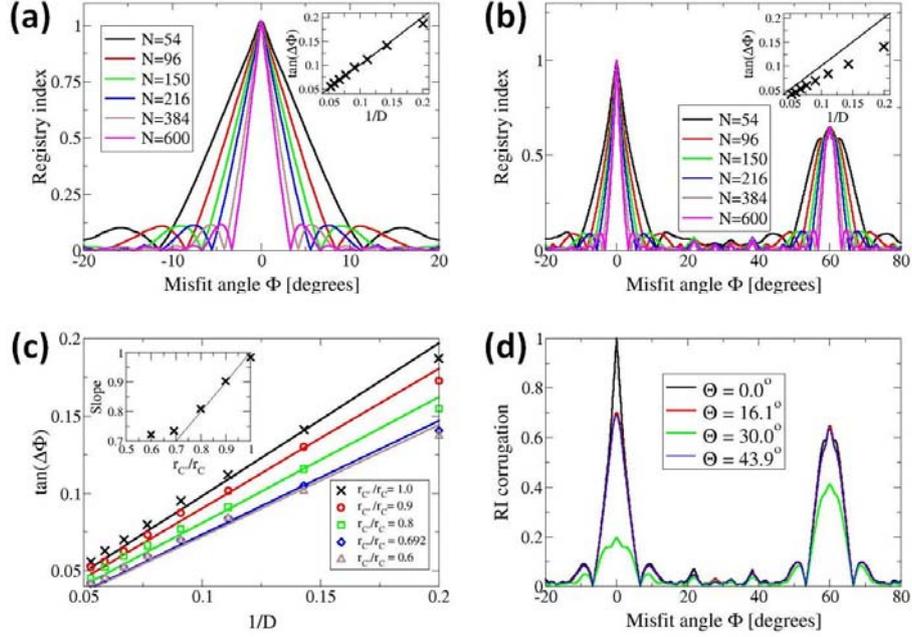

*Fig. 5: Peak shape dependence on the graphene flake size and sliding direction. **Panel (a)**: peak shape dependence on the flake size for the bilayer (symmetric) system. N stands for the number of carbon atoms in the flake. Inset: tangent of the peak width at half maximum plotted against the inverse flake diameter. **Panel (b)**: peak shape dependence on the flake size for the multilayer (asymmetric) system. Inset: tangent of the higher peak width at half maximum plotted against the inverse flake diameter. **Panel (c)**: tangent of the higher peak width at half-maximum plotted against the inverse flake diameter for various site asymmetries ($r_{C'}/r_C$) in the multilayer system. Inset: the slopes of the curves appearing in the main panel plotted against the site asymmetry. **Panel (d)**: RI corrugation as a function of misfit angle plotted for various sliding directions of the multilayer (asymmetric) system.*

Finally, we utilize the RI concept to predict the occurrence of superlubricity in *h*-BN. The model system is identical to that depicted in Fig. 1 with both graphene flake and sheet replaced by *h*-BN. For the latter, the optimal and worst stacking modes are identified as the AA' and AA configurations, respectively (see Fig. 6a, 6b). We assign a circle to each atomic site with the following radii: $r_N =$

$0.50 \cdot L_{BN}$, $r_B = 0.15 \cdot L_{BN}$,[50] where $L_{BN} = 1.44$ Å is the B-N bond length and define three projected overlap areas (Fig. 6c) between circles representing two nitrogen atoms on the two layers ($S_{NN}$) two boron atoms on the two layers ($S_{BB}$) and a boron atom on one layer with a nitrogen atom on the other layer ($S_{NB}$). Using these definitions we may construct the following expression for the RI:

$$RI_{h-BN} = \frac{(S_{NN} - S_{NN}^{AA'}) + (S_{BB} - S_{BB}^{AA'}) - (S_{NB} - S_{NB}^{AA'})}{(S_{NN}^{AA} - S_{NN}^{AA'}) + (S_{BB}^{AA} - S_{BB}^{AA'}) - (S_{NB}^{AA} - S_{NB}^{AA'})}$$

Which obtains a maximum value (*RI*=1) at the worst (AA) stacking mode and a minimum value (*RI*=0) at the optimal (AA') staking mode. Here, $S_{XY}^{AA}$ and $S_{XY}^{AA'}$ are the projected overlap areas calculated at the AA and AA' stacking modes, respectively, between a circle assigned to atom *X* on one layer and a circle assigned to atom *Y* on the other layer and summed over all atomic positions of the flake and the surface. This definition was recently shown to give excellent agreement with sliding energy landscapes obtained using advanced density functional theory calculations for *h*-BN as well as for double-walled boron nitride nanotubes.[50,94]

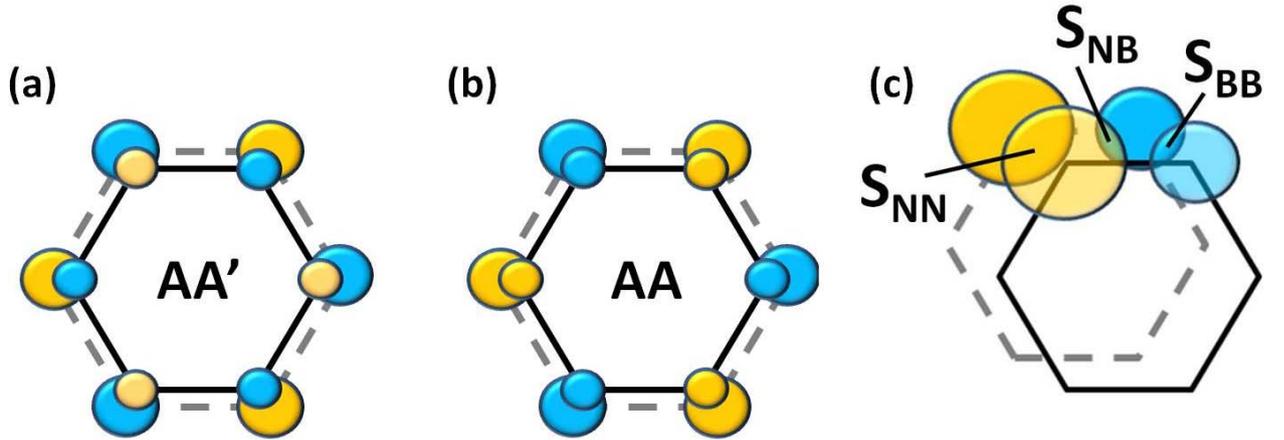

*Fig. 6: Optimal (AA', panel a) and worst (AA, panel b) staking modes of h-BN and the definition of the projected overlap area between circles assigned to atomic positions in the upper layer (transparent circles, full lines) and their lower layer counterparts (opaque circles, dashed lines). Blue and yellow circles represent boron and nitrogen atoms, respectively.*

We follow the same procedure described above where the *h*-BN flake is rotated and shifted along the underlying *h*-BN layer at a given sliding direction while recording the RI variations. We then plot the RI corrugation (maximum amplitude of the RI variations) along each linear path as a function of the misfit angle and sliding direction. Fig. 7 presents the dependence of the RI corrugation on the size of the hexagonal flake and the sliding direction. The *h*-BN system presents a pattern very similar to that obtained for graphene (Fig. 5b and 5d) where at $0^0$ and $60^0$ high friction is obtained and at intermediate angles superlubricity occurs (Fig. 7a). Despite the fact that the circles are asymmetric, when sliding along the armchair axis of the infinite *h*-BN layer the two peaks are almost identical. This results from the different definition of the registry index which stems from the different optimal and worst stacking modes of graphene and *h*-BN. A linear relation between $tan(\Delta\Phi)$ and the inverse flake diameter is achieved (inset of Fig. 7a) with a slope slightly larger than 1 ($\alpha \gtrsim 1$). Upon altering the puling direction the peak symmetry breaks and a picture very similar to that obtained for the multilayer graphene system is obtained (Fig. 7b).

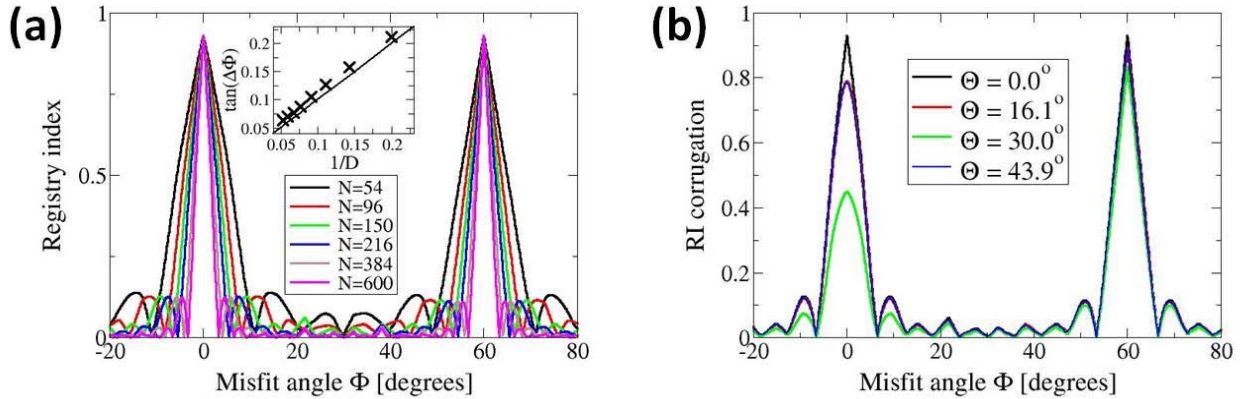

*Fig. 7: Peak shape dependence on the h-BN flake size and sliding direction. **Panel (a)**: peak shape dependence on the flake size for the bilayer system. N stands for the number of atoms in the flake. Inset: tangent of the peak width at half maximum plotted against the inverse flake diameter. **Panel (b)**: RI corrugation as a function of misfit angle plotted for various sliding directions for the bilayer h-BN system.*

To summarize, it was shown that the registry index concept provides a simple, intuitive, and extremely computationally efficient alternative for the study of tribological properties of layered materials in the regime of wearless friction. It offers a direct relation between measurable tribological phenomena, such as the occurrence of superlubricity, and geometrical parameters of the underlying interface such as the degree of lattice commensurability. In graphite the RI model was shown to be able to capture, down to fine details, the experimentally measured frictional behavior of a hexagonal graphene flake sliding on-top of the surface of graphite. It was further predicted that superlubricity is expected to occur in *h*-BN as well with tribological characteristics very similar to those observed for graphene. It should be noted that while our model is purely geometric and cannot simulate dynamical processes involving energy loss, which are of fundamental importance in studying friction, for the special case of wearless friction at the nanometer scale a direct relation between the measured kinetic friction and the corrugated sliding energy/RI landscape is obtained. Possible extensions of the RI concept towards treating hybrid layered structures such as graphene on *h*-BN, more complex layered materials such as metal dichalcogenides, curved geometries,[94] and interfaces between bulk materials are currently under study.

ACKNOWLEDGMENT: I would like to thank Prof. Joost W.M. Frenken and Prof. Martin Dienwiebel for providing me with the experimental data and granting me the rights to reproduce them in this manuscript. Many thanks to Prof. Michael Urbakh for providing me with the idea of demonstrating commensurate/incommensurate sliding using the egg-box foam model and for granting me his approval to use it within this manuscript. I would further like to thank my best friend, partner, love, and wife Adi Zalckvar Hod and my friend and colleague Dr. Yael Roichman for their help in producing Fig. 2. This work was supported by the Israel Science Foundation under grant No. 1313/08, the Center for Nanoscience and Nanotechnology at Tel Aviv University, and the Lise Meitner-Minerva Center for Computational Quantum Chemistry. The research leading to these results has received funding from the European Community's Seventh Framework Programme FP7/2007-2013 under grant agreement No. 249225.

(1) Savage, R. H.: Graphite Lubrication. *Journal of Applied Physics* **1948**, *19*, 1-10.
(2) Arnell, R. D.; Teer, D. G.: Lattice Parameters of Graphite in relation to Friction and Wear. *Nature* **1968**, *218*, 1155-1156.
(3) Bragg, W. H.: *An Introduction to Crystal Analysis*; G. Bell and sons, ltd.: London, 1928.
(4) Hyunsoo, L.; et al.: Comparison of frictional forces on graphene and graphite. *Nanotechnology* **2009**, *20*, 325701.
(5) Yen, B. K.; Schwickert, B. E.; Toney, M. F.: Origin of low-friction behavior in graphite investigated by surface x-ray diffraction. *Applied Physics Letters* **2004**, *84*, 4702-4704.
(6) Cohen, S. R.; Rapoport, L.; Ponomarev, E. A.; Cohen, H.; Tsirlina, T.; Tenne, R.; Lévy-Clément, C.: The tribological behavior of type II textured MX2 (M=Mo, W; X=S, Se) films. *Thin Solid Films* **1998**, *324*, 190-197.
(7) Rapoport, L.; Fleischer, N.; Tenne, R.: Fullerene-like WS2 Nanoparticles: Superior Lubricants for Harsh Conditions. *Advanced Materials* **2003**, *15*, 651-655.
(8) Miura, K.; Kamiya, S.: Observation of the Amontons-Coulomb law on the nanoscale: Frictional forces between MoS 2 flakes and MoS 2 surfaces. *EPL (Europhysics Letters)* **2002**, *58*, 610.
(9) Sheehan, P. E.; Lieber, C. M.: Nanotribology and Nanofabrication of MoO3 Structures by Atomic Force Microscopy. *Science* **1996**, *272*, 1158-1161.
(10) Lee, C.; Li, Q.; Kalb, W.; Liu, X.-Z.; Berger, H.; Carpick, R. W.; Hone, J.: Frictional Characteristics of Atomically Thin Sheets. *Science* **2010**, *328*, 76-80.
(11) Donnet, C.; Erdemir, A.: Historical developments and new trends in tribological and solid lubricant coatings. *Surface and Coatings Technology* **2004**, *180–181*, 76-84.
(12) Singer, I. L.; Pollock, H. M.: *Fundamentals of Friction: Macroscopic and Microscopic Processes* Kluwer Academic Publishers: Dordrecht, Netherlands, 1992.
(13) Filleter, T.; McChesney, J. L.; Bostwick, A.; Rotenberg, E.; Emtsev, K. V.; Seyller, T.; Horn, K.; Bennewitz, R.: Friction and Dissipation in Epitaxial Graphene Films. *Physical Review Letters* **2009**, *102*, 086102.
(14) Brown, S.; Musfeldt, J. L.; Mihut, I.; Betts, J. B.; Migliori, A.; Zak, A.; Tenne, R.: Bulk vs Nanoscale WS2:  Finite Size Effects and Solid-State Lubrication. *Nano Letters* **2007**, *7*, 2365-2369.
(15) Graphite on its own is not a good solid lubricant`, it requires humidity or low vapor pressure of organic additives to exhibit good lubrication characteristics.
(16) Dienwiebel, M.; Verhoeven, G. S.; Pradeep, N.; Frenken, J. W. M.; Heimberg, J. A.; Zandbergen, H. W.: Superlubricity of Graphite. *Physical Review Letters* **2004**, *92*, 126101.
(17) Dienwiebel, M.; Pradeep, N.; Verhoeven, G. S.; Zandbergen, H. W.; Frenken, J. W. M.: Model experiments of superlubricity of graphite. *Surface Science* **2005**, *576*, 197-211.
(18) Hirano, M.; Shinjo, K.: Atomistic locking and friction. *Physical Review B* **1990**, *41*, 11837.
(19) Shinjo, K.; Hirano, M.: Dynamics of friction: superlubric state. *Surface Science* **1993**, *283*, 473-478.
(20) Martin, J. M.; Donnet, C.; Le Mogne, T.; Epicier, T.: Superlubricity of molybdenum disulphide. *Physical Review B* **1993**, *48*, 10583.
(21) Hirano, M.; Shinjo, K.; Kaneko, R.; Murata, Y.: Observation of Superlubricity by Scanning Tunneling Microscopy. *Physical Review Letters* **1997**, *78*, 1448.
(22) Hirano, M.: Atomistics of friction. *Surface Science Reports* **2006**, *60*, 159-201.

(23) Tartaglino, U.; et al.: Role of surface roughness in superlubricity. *Journal of Physics: Condensed Matter* **2006**, *18*, 4143.
(24) Müser, M. H.: Theoretical Aspects of Superlubricity
Fundamentals of Friction and Wear. Gnecco, E., Meyer, E., Eds.; Springer Berlin Heidelberg, 2007; pp 177-199.
(25) Sasaki, N.; Itamura, N.; Tsuda, D.; Miura, K.: Nanomechanical Studies of Superlubricity. *Current Nanoscience* **2007**, *3*, 105-115.
(26) Sorensen, M. R.; Jacobsen, K. W.; Stoltze, P.: Simulations of atomic-scale sliding friction. *Physical Review B* **1996**, *53*, 2101.
(27) de Gennes, P.-G.: Friction between two misoriented crystalline monolayers. *Comptes Rendus Physique* **2006**, *7*, 267-271.
(28) Kim, S. H.; Asay, D. B.; Dugger, M. T.: Nanotribology and MEMS. *Nano Today* **2007**, *2*, 22-29.
(29) Bhushan, B.; Israelachvili, J. N.; Landman, U.: Nanotribology: friction, wear and lubrication at the atomic scale. *Nature* **1995**, *374*, 607-616.
(30) Braun, O. M.; Naumovets, A. G.: Nanotribology: Microscopic mechanisms of friction. *Surface Science Reports* **2006**, *60*, 79-158.
(31) Urbakh, M.; Klafter, J.; Gourdon, D.; Israelachvili, J.: The nonlinear nature of friction. *Nature* **2004**, *430*, 525-528.
(32) Gnecco, E.; et al.: Friction experiments on the nanometre scale. *Journal of Physics: Condensed Matter* **2001**, *13*, R619.
(33) Hölscher, H.; Schirmeisen, A.; Schwarz, U. D.: Principles of atomic friction: from sticking atoms to superlubric sliding. *Philosophical Transactions of the Royal Society A: Mathematical, Physical and Engineering Sciences* **2008**, *366*, 1383-1404.
(34) Frenken, J.: Nanotribology: Bringing friction to a halt. *Nat Nano* **2006**, *1*, 20-21.
(35) Cumings, J.; Zettl, A.: Low-Friction Nanoscale Linear Bearing Realized from Multiwall Carbon Nanotubes. *Science* **2000**, *289*, 602-604.
(36) Grierson, D. S.; Carpick, R. W.: Nanotribology of carbon-based materials. *Nano Today* **2007**, *2*, 12-21.
(37) Kis, A.; Jensen, K.; Aloni, S.; Mickelson, W.; Zettl, A.: Interlayer Forces and Ultralow Sliding Friction in Multiwalled Carbon Nanotubes. *Physical Review Letters* **2006**, *97*, 025501.
(38) Buldum, A.; Lu, J. P.: Atomic Scale Sliding and Rolling of Carbon Nanotubes. *Phys. Rev. Lett.* **1999**, *83*, 5050.
(39) Falvo, M. R.; Steele, J.; Taylor, R. M.; Superfine, R.: Gearlike rolling motion mediated by commensurate contact: Carbon nanotubes on HOPG. *Physical Review B* **2000**, *62*, R10665.
(40) Falvo, M. R.; Steele, J.; Taylor, R. M.; Superfine, R.: Evidence of commensurate contact and rolling motion: AFM manipulation studies of carbon nanotubes on HOPG. *Tribology Letters* **2000**, *9*, 73-76.
(41) Falvo, M. R.; Taylor Ii, R. M.; Helser, A.; Chi, V.; Brooks Jr, F. P.; Washburn, S.; Superfine, R.: Nanometre-scale rolling and sliding of carbon nanotubes. *Nature* **1999**, *397*, 236-238.
(42) Forró, L.: Beyond Gedanken Experiments. *Science* **2000**, *289*, 560-561.
(43) Shirai, Y.; Osgood, A. J.; Zhao, Y.; Kelly, K. F.; Tour, J. M.: Directional Control in Thermally Driven Single-Molecule Nanocars. *Nano Letters* **2005**, *5*, 2330-2334.
(44) Zheng, Q.; Jiang, B.; Liu, S.; Weng, Y.; Lu, L.; Xue, Q.; Zhu, J.; Jiang, Q.; Wang, S.; Peng, L.: Self-Retracting Motion of Graphite Microflakes. *Physical Review Letters* **2008**, *100*, 067205.
(45) Hirano, M.; Shinjo, K.; Kaneko, R.; Murata, Y.: Anisotropy of frictional forces in muscovite mica. *Physical Review Letters* **1991**, *67*, 2642.
(46) Ko, J. S.; Gellman, A. J.: Friction Anisotropy at Ni(100)/Ni(100) Interfaces. *Langmuir* **2000**, *16*, 8343-8351.

(47) Mancinelli, C. M.; Gellman, A. J.: Friction Anisotropy at Pd(100)/Pd(100) Interfaces. *Langmuir* **2004**, *20*, 1680-1687.

(48) Manini, N.; Braun, O. M.: Crystalline misfit-angle implications for solid sliding. *Physics Letters A* **2011**, *375*, 2946-2951.

(49) Merkle, A. P.; Marks, L. D.: Comment on "friction between incommensurate crystals". *Philosophical Magazine Letters* **2007**, *87*, 527-532.

(50) Marom, N.; Bernstein, J.; Garel, J.; Tkatchenko, A.; Joselevich, E.; Kronik, L.; Hod, O.: Stacking and Registry Effects in Layered Materials: The Case of Hexagonal Boron Nitride. *Phys. Rev. Lett.* **2010**, *105*, 046801.

(51) Kolmogorov, A. N.; Crespi, V. H.: Registry-Dependent Interlayer Potential for Graphitic Systems. *Phys. Rev. B* **2005**, *71*, 235415.

(52) Kolmogorov, A. N.; Crespi, V. H.: Smoothest Bearings: Interlayer Sliding in Multiwalled Carbon Nanotubes. *Physical Review Letters* **2000**, *85*, 4727.

(53) Liang, T.; Sawyer, W. G.; Perry, S. S.; Sinnott, S. B.; Phillpot, S. R.: First-principles determination of static potential energy surfaces for atomic friction in $MoS_{2}$ and $MoO_{3}$. *Physical Review B* **2008**, *77*, 104105.

(54) Kim, H.-J.; Kim, D.-E.: Nano-scale friction: A review. *International Journal of Precision Engineering and Manufacturing* **2009**, *10*, 141-151.

(55) Robbins, M. O.; Muser, M. H.: Computer Simulations of Friction, Lubrication and Wear. In *Modern Tribology Handbook*; Bhushan, B., Ed.; CRC Press: Boca Raton, FL, 2001; pp 717-765.

(56) Legoas, S. B.; Giro, R.; Galvão, D. S.: Molecular dynamics simulations of C60 nanobearings. *Chemical Physics Letters* **2004**, *386*, 425-429.

(57) Szlufarska, I.; Chandross, M.; Carpick, R. W.: Recent advances in single-asperity nanotribology. *Journal of Physics D: Applied Physics* **2008**, *41*, 123001.

(58) Hayashi, K.; Maeda, A.; Terayama, T.; Sakudo, N.: Molecular dynamics study of wearless friction in sub-micrometer size mechanisms and actuators based on an atomistic simplified model. *Computational Materials Science* **2000**, *17*, 356-360.

(59) Carlson, A.; Dumitrică, T.: Extended Tight-Binding Potential for Modelling Intertube Interactions in Carbon Nanotubes. *Nanotechnology* **2007**, *18*, 065706.

(60) Popov, A. M.; Lebedeva, I. V.; Knizhnik, A. A.; Lozovik, Y. E.; Potapkin, B. V.: Commensurate-incommensurate phase transition in bilayer graphene. *Physical Review B* **2011**, *84*, 045404.

(61) Onodera, T.; Morita, Y.; Suzuki, A.; Koyama, M.; Tsuboi, H.; Hatakeyama, N.; Endou, A.; Takaba, H.; Kubo, M.; Dassenoy, F.; Minfray, C.; Joly-Pottuz, L.; Martin, J.-M.; Miyamoto, A.: A Computational Chemistry Study on Friction of h-MoS2. Part I. Mechanism of Single Sheet Lubrication. *J. Phys. Chem. B* **2009**, *113*, 16526-16536.

(62) Zhong, W.; Tomanek, D.: First-principles theory of atomic-scale friction. *Physical Review Letters* **1990**, *64*, 3054.

(63) Zhang, C.: Computational Investigation on the Desensitizing Mechanism of Graphite in Explosives versus Mechanical Stimuli: Compression and Glide. *The Journal of Physical Chemistry B* **2007**, *111*, 6208-6213.

(64) Calvo-Almazan, T.; Seydel, T.; Fouquet, P.: Questions arising for future surface diffusion studies using scattering techniques—the case of benzene diffusion on graphite basal plane surfaces. *Journal of Physics: Condensed Matter* **2010**, *22*, 304014.

(65) Lebedeva, I. V.; Knizhnik, A. A.; Popov, A. M.; Lozovik, Y. E.; Potapkin, B. V.: Interlayer interaction and relative vibrations of bilayer graphene. *Physical Chemistry Chemical Physics* **2011**, *13*, 5687-5695.

(66) Cahangirov, S.; Ataca, C.; Topsakal, M.; Sahin, H.; Ciraci, S.: Frictional Figures of Merit for Single Layered Nanostructures. *Physical Review Letters* **2012**, *108*, 126103.

(67) Singer, I. L.: Friction and energy dissipation at the atomic scale: A review. *Journal of Vacuum Science and Technology A* **1994**, *12*, 2605-2616.
(68) Verhoeven, G. S.; Dienwiebel, M.; Frenken, J. W. M.: Model calculations of superlubricity of graphite. *Physical Review B* **2004**, *70*, 165418.
(69) Enrico, G.; et al.: Superlubricity of dry nanocontacts. *Journal of Physics: Condensed Matter* **2008**, *20*, 354004.
(70) Filippov, A. E.; Dienwiebel, M.; Frenken, J. W. M.; Klafter, J.; Urbakh, M.: Torque and Twist against Superlubricity. *Physical Review Letters* **2008**, *100*, 046102.
(71) Steiner, P.; Roth, R.; Gnecco, E.; Baratoff, A.; Maier, S.; Glatzel, T.; Meyer, E.: Two-dimensional simulation of superlubricity on NaCl and highly oriented pyrolytic graphite. *Physical Review B* **2009**, *79*, 045414.
(72) Wang, C.-L.; et al.: Application of Two-Dimensional Frenkel–Kontorova Model to Nanotribology. *Communications in Theoretical Physics* **2010**, *54*, 112.
(73) Miura, K.; Sasaki, N.; Kamiya, S.: Friction mechanisms of graphite from a single-atomic tip to a large-area flake tip. *Physical Review B* **2004**, *69*, 075420.
(74) de Wijn, A. S.; Fasolino, A.; Filippov, A. E.; Urbakh, M.: Low friction and rotational dynamics of crystalline flakes in solid lubrication. *EPL* **2011**, *95*, 66002.
(75) Prandtl, L.: Ein Gedankenmodell zur kinetischen Theorie der festen Körper. *ZAMM - Journal of Applied Mathematics and Mechanics / Zeitschrift für Angewandte Mathematik und Mechanik* **1928**, *8*, 85-106.
(76) Tomlinson, G. A.: CVI. A molecular theory of friction. *Philosophical Magazine Series 7* **1929**, *7*, 905-939.
(77) Frenkel, Y. I.; Kontorova, T. A.: *Zh. Eksp. Teor. Fiz.* **1938**, *8*, 89.
(78) Ershova, O. V.; Lillestolen, T. C.; Bichoutskaia, E.: Study of polycyclic aromatic hydrocarbons adsorbed on graphene using density functional theory with empirical dispersion correction. *Physical Chemistry Chemical Physics* **2010**, *12*, 6483-6491.
(79) Björk, J.; Hanke, F.; Palma, C.-A.; Samori, P.; Cecchini, M.; Persson, M.: Adsorption of Aromatic and Anti-Aromatic Systems on Graphene through π−π Stacking. *J. Phys. Chem. Lett.* **2010**, *1*, 3407-3412.
(80) Guo, Y.; Guo, W.; Chen, C.: Modifying atomic-scale friction between two graphene sheets: A molecular-force-field study. *Physical Review B* **2007**, *76*, 155429.
(81) Itamura, N.; Miura, K.; Sasaki, N.: Simulation of Scan-Directional Dependence of Superlubricity of C60 Molecular Bearings and Graphite. *Japanese Journal of Applied Physics* **2009**, *48*, 060207.
(82) Khomenko, A. V.; Prodanov, N. V.: Molecular dynamics of cleavage and flake formation during the interaction of a graphite surface with a rigid nanoasperity. *Carbon* **2010**, *48*, 1234-1243.
(83) Onodera, T.; Morita, Y.; Nagumo, R.; Miura, R.; Suzuki, A.; Tsuboi, H.; Hatakeyama, N.; Endou, A.; Takaba, H.; Dassenoy, F.; Minfray, C.; Joly-Pottuz, L.; Kubo, M.; Martin, J.-M.; Miyamoto, A.: A Computational Chemistry Study on Friction of h-MoS2. Part II. Friction Anisotropy. *J. Phys. Chem. B* **2010**, *114*, 15832-15838.
(84) de Wijn, A. S.; Fusco, C.; Fasolino, A.: Stability of superlubric sliding on graphite. *Physical Review E* **2010**, *81*, 046105.
(85) Kim, W. K.; Falk, M. L.: Atomic-scale simulations on the sliding of incommensurate surfaces: The breakdown of superlubricity. *Physical Review B* **2009**, *80*, 235428.
(86) Bonelli, F.; Manini, N.; Cadelano, E.; Colombo, L.: Atomistic simulations of the sliding friction of graphene flakes. *Eur. Phys. J. B* **2009**, *70*, 449-459.
(87) Fouquet, P.; Johnson, M. R.; Hedgeland, H.; Jardine, A. P.; Ellis, J.; Allison, W.: Molecular dynamics simulations of the diffusion of benzene sub-monolayer films on graphite basal plane surfaces. *Carbon* **2009**, *47*, 2627-2639.

(88) Guerra, R.; Tartaglino, U.; Vanossi, A.; Tosatti, E.: Ballistic nanofriction. *Nat Mater* **2010**, *9*, 634-637.

(89) Liang, X.; et al.: Vanishing stick–slip friction in few-layer graphenes: the thickness effect. *Nanotechnology* **2011**, *22*, 285708.

(90) Liang, T.; Phillpot, S. R.; Sinnott, S. B.: Parametrization of a reactive many-body potential for Mo–S systems. *Physical Review B* **2009**, *79*, 245110.

(91) Mylvaganam, K.; Zhang, L. C.: Nano-Friction of Some Carbon Allotropes. *Journal of Computational and Theoretical Nanoscience* **2010**, *7*, 2199-2202.

(92) Lebedeva, I. V.; Knizhnik, A. A.; Popov, A. M.; Ershova, O. V.; Lozovik, Y. E.; Potapkin, B. V.: Fast diffusion of a graphene flake on a graphene layer. *Physical Review B* **2010**, *82*, 155460.

(93) Lebedeva, I. V.; Knizhnik, A. A.; Popov, A. M.; Ershova, O. V.; Lozovik, Y. E.; Potapkin, B. V.: Diffusion and drift of graphene flake on graphite surface. *Journal of Chemical Physics* **2011**, *134*, 104505.

(94) Hod, O.: Quantifying the Stacking Registry Matching in Layered Materials. *Isr. J. Chem.* **2010**, *50*, 506-514.